# Magneto-Electric Effect for Multiferroic Thin Film by Monte Carlo Simulation

## Zidong Wang (王子东), Malcolm J. Grimson

Department of Physics, University of Auckland, Auckland 1010, New Zealand
*corresponding author, E-mail: Zidong.Wang@auckland.ac.nz

## Abstract

Magneto-electric effect in a multiferroic heterostructure film, i.e. a coupled ferromagnetic-ferroelectric thin film, has been investigated through the use of the Metropolis algorithm in Monte Carlo simulations. A classical Heisenberg model describes the energy stored in the ferromagnetic film, and we use a pseudo-spin model with a transverse Ising Hamiltonian to characterise the energy of electric dipoles in the ferroelectric film. The purpose of this article is to demonstrate the dynamic response of polarisation is driven by an external magnetic field, when there is a linear magneto-electric coupling at the interface between the ferromagnetic and ferroelectric components.

## 1. Introduction

Recently, the magneto-electric (ME) effect in thin multiferroic films has been investigated theoretically [1-9] and experimentally [10,11]. Our understanding the nature of the ME effect has two origins. One is the physical mechanism coupling the magnetic and electric response by elastic interactions [12-14]. As *Fig. 1* shows, a magnetised material produces mechanical strain due to the magnetostrictive deformation. Then the strain is passed to the adjacent electric material, resulting in a polarisation, due to the shape change by the stress. Thus, a magnetic-mechanical effect in the ferromagnet (FM) and a mechanical-electric effect in the ferroelectric (FE) constitute the ME effect in the multiferroic materials. Another origin is related to the electrostatic screening and has been studied by T. Y. Cai, *et.al.* [15] and C. L. Jia, *et.al.* [4]. Generally, the FM part is a normal magnetic metal (e.g. iron, nickel and cobalt) or their compounds [2,4,10,15]; whereas the FE part is, for example, barium titanate ($BaTiO_3$), lead titanate ($PbTiO_3$), or lead zirconate titanate (PZT) [2,4,10,15,16].

In this work, to study the mechanism of the ME effect, we introduce a FM/FE coupled thin film model with magnetic spins in the FM part and electric dipoles located in the FE part. The system is driven by a dynamic magnetic field. The mean responses of the magnetic spins (magnetisation) and the electric dipoles (polarisation) have been obtained, by using the Metropolis algorithm in a kinetic Monte Carlo simulation. The details are given in section 2. The numerical results are presented in section 3. An analysis about the ME effect with distinct ME couplings is drawn in section 4. A summary is in section 5.

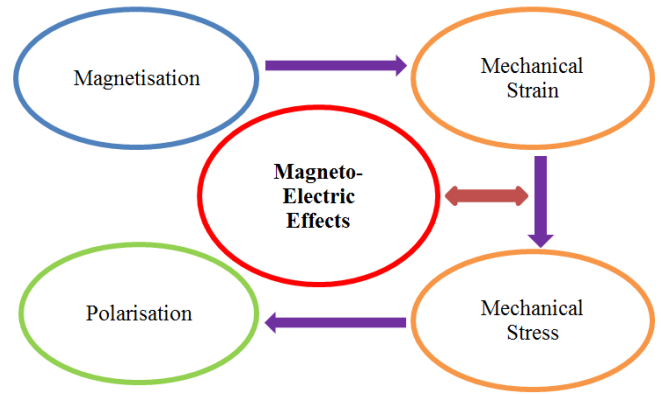

*Fig. 1. (Colour online) Principle of the ME effect, driven by the magnetic material.*

## 2. Model and Method

The composite material incorporates FM and FE films, is called the multiferroic heterostructure. The dynamic ME effect is defined as the induced polarisation of a multiferroic material in a time-dependent external magnetic field. Our numerical model clearly exhibits the aforementioned phenomenon. In *Fig. 2*, a sketch of coupled FM and FE films in a 2-D lattice. The *blue arrows* represent the magnetic spins, and the *red arrows* represent the locations of electric dipole. Each magnetic spin and electric dipole has three degrees of freedom. Thus, a simple electric pseudo-spin model [3,5,17-19] describes the electric dipoles as the pseudo-spins. The ME effect occurs at the interface, where $g$ represents the ME coupling (*yellow wall*). A time-dependent external magnetic field $B(t)$ is applied parallel to the interface (*violet arrow*).

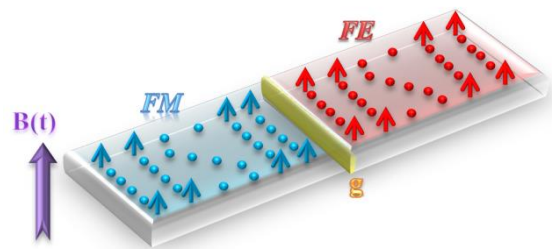

*Fig. 2. (Colour online) Schematic of a multiferroic heterostructure thin film, the blue arrows represent the magnetic spins, and the red arrows represent the electric*

*pseudo-spins. An external magnetic field is applied (violet arrow). The interface is represented by the yellow wall.*

The system under consideration is operating with a classical anisotropic Heisenberg model and a transverse Ising model. The following Hamiltonians have been employed for each FM layer, $\mathcal{H}_{FM}$ [20], and each FE layer, $\mathcal{H}_{FE}$ [21]:

$$\mathcal{H}_{FM} = -J_{FM} \sum_{\langle i,j \rangle}^{N_L} \left( S_i^x S_j^x + S_i^y S_j^y + S_i^z S_j^z \right) - K \sum_i^{N_L} \left( S_i^z \right)^2 \quad (1)$$

and,

$$\mathcal{H}_{FE} = -\Omega \sum_i^{N_L} P_i^x - J_{FE} \sum_{\langle i,j \rangle}^{N_L} \left( P_i^z P_j^z \right) \quad (2)$$

where $S_i^{x,y,z}$ and $P_i^{x,y,z}$ denote the magnetic spin and the electric pseudo-spin components at film site $i$ with the unit size, i.e., $\left| S_i^{x,y,z} \right| = \left| P_i^{x,y,z} \right| = 1$. The notation $\langle i,j \rangle$ characterises that the sum is restricted to nearest-neighbour pairs of spins, each pair being counted only once. $J_{FM}$ and $J_{FE}$ are the nearest-neighbour exchange interaction strengths for the FM film and FE film, respectively. Since an electric dipole is a separation of positive and negative charges, a measure of this separation gives the magnitude of the electric dipole moment [5,17-19]. It is a scalar. Thus, the exchange interaction energy in *Eq. (2)* only contains z-component polarisations. K is the uniaxial magnetic anisotropy coefficient in the z-direction. $\Omega$ is the x-direction transverse field to the electric pseudo-spins. $L_{FM}$ and $L_{FE}$ represent the number of layers in the FM and FE films, respectively. $L = L_{FM} + L_{FE}$ is the total number of layers in this sample. Each layer contains a fixed $N_L$ number of spins/pseudo-spins, and there are $N = N_L \times L$ number of samples in total system. The Zeeman energy $\mathcal{H}_{ext}$, between magnetic spins and external magnetic field $B(t)$, applied in the z-direction are described by:

$$\mathcal{H}_{ext} = -B(t) \sum_i^{N_L} S_i^z \quad (3)$$

The ME effect at the interface has been considered as a result of a linear coupling [1-3,5]. *Eq. (4)* represents the interface energy for the magnetic spins in the last FM layer $S_{i,L_{FM}}$ and the electric pseudo-spins in the first FE layer $P_{i,1}$.

$$\mathcal{H}_{ME} = -g \sum_i^{N_L} \left( S_{i,L_{FM}}^z P_{i,1}^z \right) \quad (4)$$

Thus, the sum of the energies gives the total energy in the system as,

$$\mathcal{H} = \mathcal{H}_{ME} + \sum^{L_{FM}} (\mathcal{H}_{FM} + \mathcal{H}_{ext}) + \sum^{L_{FE}} (\mathcal{H}_{FE}) \quad (5)$$

Using the Metropolis algorithm in a kinetic Monte Carlo simulation [22-25], the formula

$$P_{RT} = \exp(-\beta \Delta E) \quad (6)$$

is used to calculate the transition probability. Where $\Delta E$ is the change of the total energy, $\beta = (k_B T)^{-1}$ is the 'inverse temperature', and $k_B$ is the Boltzmann's constant. The entire process of the simulation method is shown in *Fig. 3*.

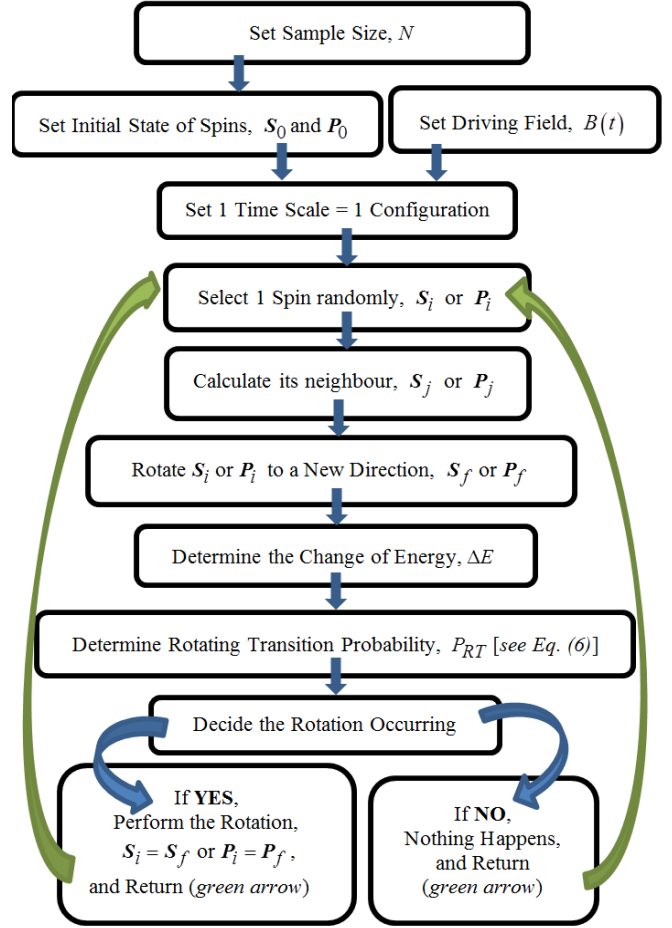

*Fig. 3. (Colour online) This figure illustrates the steps of the Metropolis algorithm in a kinetic Monte Carlo simulation.*

### 3. Results

To demonstrate the dynamic behaviour of magnetic and electric responses, we set three first layers in the FM film, i.e. $L_{FM} = 3$, and six next layers in the FE film, i.e. $L_{FE} = 6$. More layers in the FE part allows focus on the ME effect in the paper. The FM part is directly driven by the magnetic field. Thus we only need to consider the different number of the neighbours in the surface, bulk, and interface. The number of spins/pseudo-spins is set at $N_L = 1000$ in each layer. Same value of the nearest-neighbour exchange interaction couplings $J_{FM} = J_{FE} = 1$. The coefficient of anisotropy is K = 0.1 in the FM. The strength of the transverse field in the FE is $\Omega = 0.1$. The time-dependent magnetic driving field is set in a sinusoidal type $B(t) = B_0 sin(\omega t)$, with fixed amplitude of $B_0 = 10$, and the switching frequency is $\omega = \pi/900000$. In our Monte Carlo approach, the 'inverse temperature' is normalised as $\beta=1$. Simulations are performed for up to 200 Monte Carlo configurations per spin in one period [24-26]. Thus, the total scan of three periods is $5.4\times 10^6$ configurations to ensure equilibration of the system. The unit of time is equal to one configuration as shown in results. Free boundary conditions have been applied in the first layer in the FM film and the last layer in the FE film. Periodic boundary conditions are used in each layer for the first spin and the last spin. The initial states of $S_0$ and $P_0$ have been set at random.

The simulation results show the *z*-component of mean magnetic response (magnetisation), $S_z$ and electric response (polarisation), $P_z$ per spin in each layer. *Fig. 4* shows the different dynamic responses in each layer for a ME coupling $g=1$, with three periods. In *Figs. 4(a)*-(c), the magnetisations in the FM layers show fully saturated responses, since they are directly driven by the external field. In the FE film, the energy transfer is only executed by nearest-neighbour interactions. Thus, the amplitudes of the polarisation taper off quickly, with a longer delay time for the further layers, are shown in *Figs. 4(d)-(i)*.

From another perspective, the hysteresis loop in each layer is shown in *Fig. 5*, with same data used in *Fig. 4*. Normal symmetric hysteresis loops are obtained in the FM film with respect to the saturated responses (*blue loops* in *Fig. 5*). Since the electric polarisation (*red loops* in *Fig. 5*), are indirectly driven by the magnetic driving field, changes its characteristic from symmetric to asymmetric with penetration of the FE layers. The hysteresis curves are asymmetric loops in *Figs. 5(e)-(i)*. This up-shifting results from the choice of initial conditions. The area contained the loop shows decaying behaviour with the further layers into the FE film. Moreover, the hysteresis curves in *Figs. 5(g)-(i)* exhibit slightly asymmetry, i.e. the left-hand side is higher than the right-hand side. This is a result of the loop lying in the upper half of the figure due to the initial conditions. Detailed explanations are given in section 4.

Two 3-D plots of the time-dependent responses are shown in *Fig. 6*, for two ME couplings $g=1$ and $g=0.3$. The multiple colours in this figure characterise the *z*-component magnitudes of the mean magnetisation/polarisation at each layer. The dynamic response for the ME coupling $g=1$, from the data in *Figs. 4 & 5*, is given in the *top panel* of *Fig 6*. Comparison with the results in the *bottom panel* of *Fig. 6* for a similar system with a different ME coupling, $g=0.3$. This shows that the ME effect is controlled by the ME coupling. In the *bottom panel* of *Fig. 6*, a weak ME coupling gives dramatic decay of the electric polarisation after the interface.

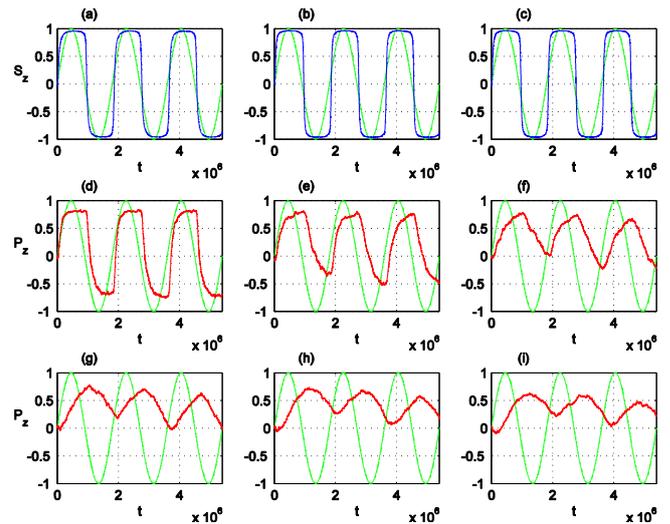

*Fig. 4. (Colour online) The mean dynamic z-component responses in each layer, to a reduced magnetic driving field (green curves), with a ME coupling $g=1$. (a)-(c) show the magnetic responses (magnetisations), $S_z$ in the three first FM layers; (d)-(i) show the electric responses (polarisations), $P_z$ in the six next FE layers.*

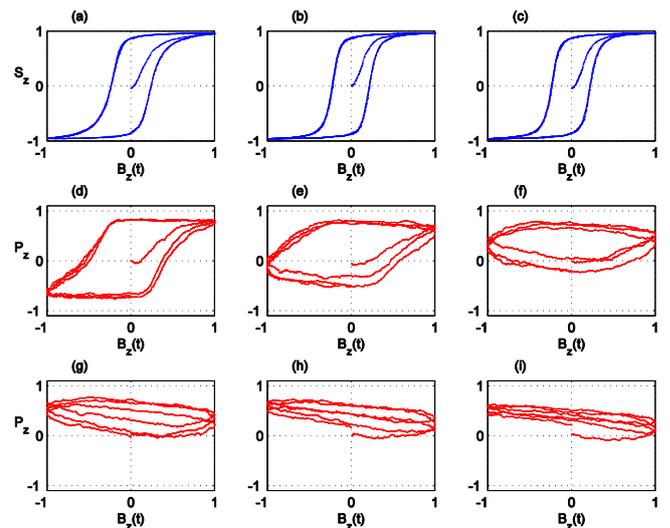

*Fig. 5. (Colour online) The hysteresis loops in different layers which shows the mean magnetisations (blue loops) and polarisations (red loops) in response to the magnetic driving field. The same data from Figure 4 is used.*

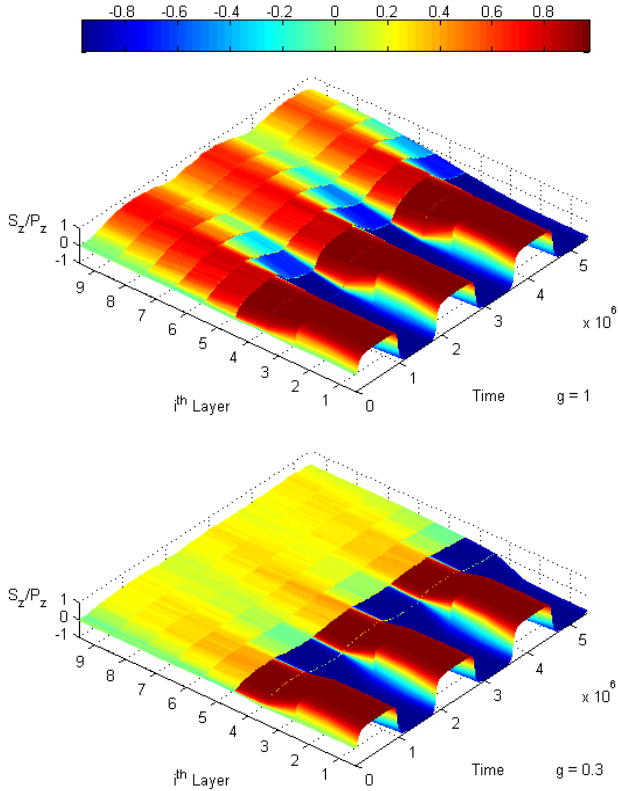

*Fig. 6. (Colour online) The 3-D plots show the mean z-component response in each layer, in respect of time. The top panel for a ME coupling, $g = 1$, and the bottom panel for $g = 0.3$.*

### 4. Effects of the Magneto-Electric Coupling

In order to study the dependence on ME effect of the multiferroic heterostructure film, simulations have been performed to determine the responses of magnetisation and polarisation for different values of the ME coupling $g$. To investigate the continuous decay behaviour of the polarisation response in each FE layer, a close inspection of the maximal magnetisation/polarisation in each layer is presented in *Fig. 7* for $0.2 \leq g \leq 1$. For the sake of clarity, each result is obtained for the system with a random initial state. The results in *Fig. 7* can be compared directly by different ME couplings. Since the FM film is driven by the driving field directly, the maximal magnetisations in first three layers are independent on the value of $g$. However, the maximal polarisations in the FE film (*layers 4-9* in *Fig. 7*), shows an exponential-like decay. A faster decay can be observed for small value of coupling $g$, due to the energy transfer being limited by the interface. As mentioned before, the ME effect decreases as the ME coupling $g$ decreases.

Earlier work has identified the delay behaviour of polarisations in the FE film. In *Fig. 8*, the *layers 4-9* show a linear-like result in the FE layers, with seven ME couplings from $0.2 \leq g \leq 1$. The linear character of the curves indicates that the speeds of the interaction energy transfer have the same rate. The only exception is for the edge layer (*layers 9* in *Fig. 8*) and it is due to the free boundary condition. *Fig. 8* further shows the curves with different colours are mixed up. This means that the time of the peak response is not related to the values of $g$. In a nutshell, the delay behaviour in the FE film is independent of the ME effect.

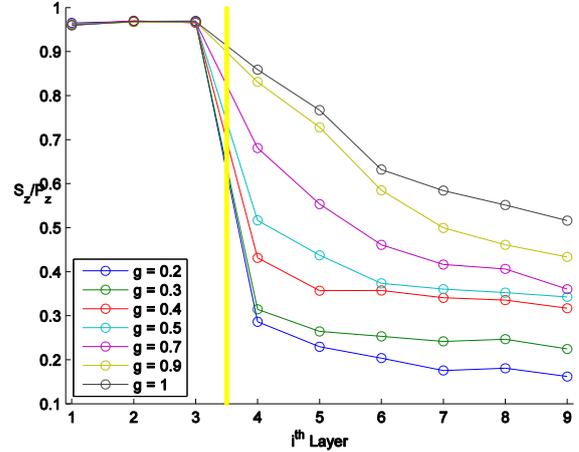

*Fig. 7. (Colour online) This figure illustrates the maximal responses in each layer, with different values of ME coupling in a range of $0.2 \leq g \leq 1$. The interface indicates in a yellow line. The lines through those points are only guide to the eye.*

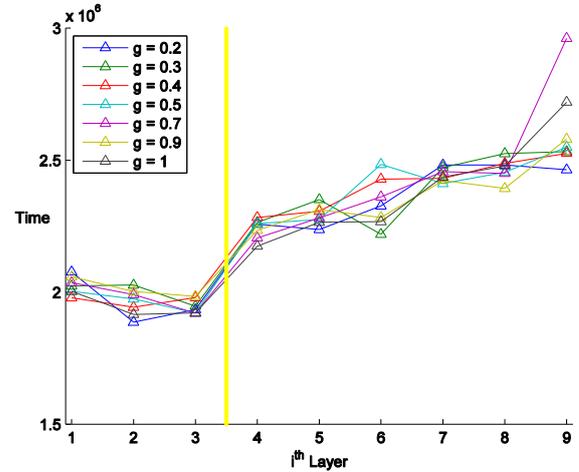

*Fig. 8. (Colour online) This figure illustrates the exact time in the second period, when the maxima responses occurred in each layer. Different values of the ME coupling depicted with individual colours, in a range of $0.2 \leq g \leq 1$. The interface indicates in a yellow line. The lines through those points are only guide to the eye.*

### 5. Conclusions

In this paper, the ME effect in a 2-D multiferroic heterostructure thin film has been demonstrated by the Metropolis algorithm in a kinetic Monte Carlo method. As a proof of concept, the polarisation was distinctly controlled by the external magnetic field as indicated by experiments. The ME coupling, $g$ is clearly seen to have a crucial role in controlling the energy transfer between the interface of the FM and FE film. The different values of the ME coupling $g$ determine the magnitude of the response in system, which exhibits an exponential decay of maximal response in each FE layer. But, the relaxation time is not changed by varying the values of coupling $g$. As an aside, the Monte Carlo approach can provide a direct temperature control by its transition probability. But converting the number of the Monte Carlo configurations to the real time has to be based on an assumption. However, the spin dynamics approach [*1-3,5*] and the Landau phenomenological theory [*1,2,27*] can easily solve this problem. Additionally, this system can also be driven by an external electric field, and shows a magnetic response in the magnetic material as well [*1,3,5*].

## Acknowledgements

The author gratefully acknowledges Zhao BingJin (赵秉金) and Wang YuHua (王玉华) for financial support.